\documentclass[aip,jcp,preprint,nofootinbib,superscriptaddress,floatfix]{revtex4-1}
\usepackage{graphicx}
\usepackage{xcolor}
\usepackage{amsmath}
\newcommand{\I}{{\mathrm{i}}}
\newcommand{\dd}{{\mathrm{d}}}
\renewcommand\Re{\operatorname{Re}}

\begin{document}
%
%
\title{On transition rates in surface hopping}
\author{J.M. Escart\'in}
\affiliation{Laboratoire de Physique Th\'{e}orique-IRSAMC, CNRS, Universit\'{e} Paul Sabatier, F-31062 Toulouse Cedex, France} 
\author{P. Romaniello}
\affiliation{Laboratoire de Physique Th\'{e}orique-IRSAMC, CNRS, Universit\'{e} Paul Sabatier, F-31062 Toulouse Cedex, France} 
\affiliation{European Theoretical Spectroscopy Facility (ETSF)}
\author{L. Stella}
\affiliation{University of the Basque Country UPV/EHU, Nano-Bio Spectroscopy Group, Avenida de Tolosa 72, 20018 San
Sebasti\'an, Spain}
\affiliation{
Department of Physics and Astronomy and London Centre for Nanotechnology, University College London, Gower Street, London WC1E 6BT, United  Kingdom}
\affiliation{European Theoretical Spectroscopy Facility (ETSF)}
\author{P.-G. Reinhard}
\affiliation{Institut f\"ur Theoretische Physik II, Universit\"at Erlangen-N\"urnberg, D-91058 Erlangen, Germany}
\author{E. Suraud}
\affiliation{Laboratoire de Physique Th\'{e}orique-IRSAMC, CNRS, Universit\'{e} Paul Sabatier, F-31062 Toulouse Cedex, France} 
\begin{abstract}
Trajectory surface hopping (TSH) is one of the most widely used
quantum-classical algorithms for nonadiabatic molecular dynamics.
Despite its empirical effectiveness and popularity, a rigorous
derivation of TSH as the classical limit of a combined quantum
electron-nuclear dynamics is still missing.  In this work we aim
  to elucidate the theoretical basis for the widely used hopping
  rules. Naturally, we concentrate thereby on the formal aspects of the
  TSH.
  
Using a Gaussian wave packet limit, we derive
the transition rates governing the hopping process at a simple avoided level crossing. In this derivation, which gives insight into the physics underlying the hopping process, some essential features of the standard TSH algorithm are retrieved, 
namely i) non-zero electronic transition rate (``hopping probability'') at avoided crossings; ii) rescaling of the nuclear velocities to conserve total energy; iii) electronic transition rates linear in the nonadiabatic coupling vectors. The well-known Landau-Zener model is then used for illustration. 
\end{abstract}
\maketitle 
\section{Introduction}
 Practical molecular modeling is based on the Born-Oppenheimer 
approximation (BOA), which allows one to decouple the fast electronic motion from the usually slower nuclear motion 
by introducing i) the (adiabatic) potential energy surfaces (PESs), provided by the electrons in a specific eigenstate, and ii) the
nonadiabatic couplings (NACs).\cite{Born-Huang} When the latter are neglected, nuclei move on a single PES. Despite the success of the BOA, there are many physical situations, such as photo-reaction, electron transfer,
or any form of non-radiative electronic relaxation, which involve more than one PES.\cite{Wodtke} In these cases one has to take into account the coupling between various PESs.

One of the most widely applied techniques to treat nonadiabatic
effects in molecular dynamics is the trajectory surface hopping (TSH),
with its several variants.%
\cite{Tully_71,Tully_90,Tully_94,Webster_94,Tully_98,Prezhdo_97,Wong_02,Wong_02_2,Schwartz_05,Subotnik_11,Subotnik_11_2,Subotnik_11_3}
The main idea behind this technique is that, while the electronic wave function is propagated coherently, the
force field felt by the nuclei varies in a discontinuous, stochastic way --- the nuclei move along a single
adiabatic PES, selected according to the electronic population of the
corresponding state; time changes in the electronic populations can result in a
sudden \textit{hop} to another adiabatic energy surface. In order to
conserve the total energy after each hop, the nuclear velocities are
rescaled.  This leads to a discontinuity in the nuclear
  velocities which, however, is generally small since the hops are
more likely to occur between PESs which are close in energy. Yet,
energy conservation is obtained in a rather \textit{ad hoc} way and,
although it is common practice to rescale the velocities along the
direction of the nonadiabatic coupling vectors (which couple
different adiabatic states),\cite{Tully_90,Truhlar_00}
in principle other choices are possible.\cite{Truhlar_00,Schwartz_05,Subotnik_11} As a consequence, most surface hopping algorithms are justified on an empirical basis,
by direct comparison with exact analytical results in model systems or experimental data. A further issue concerns the loss of 
electron-nuclear coherence in the course of the dynamics. This aspect is directly related to the 
 scaling of the transition probability with respect to the nonadiabatic couplings. It is 
 traditionally linear in standard TSH,\cite{Tully_90} but it was recently shown to be incorrect for
some model cases,\cite{Subotnik_11} the effect being directly traced back to the treatment of decoherence over time.

In this work emphasis is put on the formal analysis in order to
  provide a basis for better understanding surface hopping techniques,
  rather than concentrating on numerical aspects. This aim is similar
  in spirit to that of previous works,
  \cite{Herman_JCP84,Herman_JCP95,Herman_JCP06,Tully_98} but in a
  simpler framework.
The paper is organized as follows. In Sec.\ \ref{sec:Basics} we briefly introduce the formalism which will be used 
in the rest of the paper. In Sec.\ \ref{sec:Tully} we derive the equations governing the electronic transition rates at an avoided crossing by using a Gaussian wave packet limit for the nuclear wave function. The equations directly display the physics behind the TSH, 
i.e., the physics governing a hop at an avoided crossing followed by rescaling of the nuclear velocities in order to conserve  total energy. We find that the physical
source of the velocity rescaling is related to the speed of variation of the NACs
and that the rescaling only affects the nuclear
velocity components parallel to the NAC vectors.
We further discuss some general consequences of our theoretical approach and its relevance in the context of practical
nonadiabatic molecular dynamics.  In Sec\ \ref{sec:Illustration} the electronic transition rates are explored by using the Landau-Zener model as a paradigmatic test case. 
We finally draw our conclusions and future perspectives.

\section{Basics\label{sec:Basics}}
We consider a quantum mechanical system of $n$ electrons and $N$ nuclei 
with the total Hamiltonian $\hat{H} =\hat{ T}+ \hat{H}^\mathrm{el}$ being the sum of the kinetic energy of the nuclei,
$\hat{T}=-{{\mathbf{\nabla}}^2_{\mathbf{R}}}/{2M}$, and the electronic Hamiltonian, $\hat{ H}^\mathrm{el}$,
which contains the kinetic energy of the electrons, the electron-electron potential, the electron-nuclear coupling, 
and the nucleus-nucleus potential. To keep notations simple, throughout this paper we denote the set of electronic coordinates by $\mathbf{r}$ 
and the nuclear ones by $\mathbf{R}$; moreover we consider all nuclei having the same mass $M$ and we use atomic units.

The nuclei are much heavier than the electrons and thus it is a natural first approximation to
consider them as classical particles, having at all times well defined
positions $\mathbf{R}$ and momenta  $\mathbf{P}$.
This suggests an adiabatic procedure where the electronic problem is solved for nuclei momentarily clamped to fixed positions in space: $\hat{H}^{\mathrm{el}}(\mathbf{R})
\phi_i(\mathbf{r};\mathbf{R})=E_i(\mathbf{R})\phi_i(\mathbf{r};\mathbf{R})$. The (adiabatic) electronic eigenfunctions $\{\phi_i(\mathbf{r};\mathbf{R}) \}$ depend parametrically on the atomic
positions and form a complete and orthonormal set. They can be used as a basis to expand the total
wave function of the system as 
\begin{equation}
\Xi(\mathbf{r},\mathbf{R};t)=\sum_j\chi_j(\mathbf{R};t)\phi_j(\mathbf{r};\mathbf{R}),
\label{eqn:expansion}
\end{equation}
 and solve the time-dependent Schr\"odinger equation
 \begin{equation}
\I\partial_t\Xi(\mathbf{r},\mathbf{R};t) = \hat{H}\Xi(\mathbf{r},\mathbf{R};t).
\label{eqn:SE}
\end{equation}
The expansion coefficients $\chi_j(\mathbf{R};t)$, which depend on the nuclear positions, 
will be identified as nuclear wave packets. Such wave packets are neither orthogonal nor normalized.
In fact, the integral $\int \dd^{3N}\mathbf{R}\, |\chi_j(\mathbf{R};t)|^2=||\chi_j(t)||^2$ gives 
the instantaneous electronic population of the $j^{th}$ quantum state. 
By inserting expansion (\ref{eqn:expansion}) for the total wave function into the Schr\"odinger equation (\ref{eqn:SE})
and projecting out the resulting equation on the electronic state $\phi_i$,
one obtains
\begin{eqnarray}
 \I\frac{\partial}{\partial t}\chi_i(\mathbf{R};t)
  &=&
  \left[\hat{T}+E_i(\mathbf{R})\right]\chi_i(\mathbf{R};t)
\nonumber\\
  & &
  - \I \sum_j\mathbf{D}_{ij}(\mathbf{R})\cdot\hat{\mathbf{P}}\chi_j(\mathbf{R};t).
\label{Eqn:nuclear_WF_adiabatic}
\end{eqnarray}
with 
\begin{equation}\label{eqn:nac}
\mathbf{D}_{ij}(\mathbf{R}) = \frac{1}{M}\langle\phi_i(\mathbf{r};\mathbf{R})|\nabla_{\mathbf{R}}|\phi_j(\mathbf{r};\mathbf{R})\rangle
\end{equation}
 being the NAC vectors and $\hat{\mathbf{P}}
= -\I\nabla_{\mathbf{R}}$ the momentum operator for
the nuclei, and  
where we used the notation $\langle\phi_i(\mathbf{r};\mathbf{R})|\hat{O}|\phi_j(\mathbf{r};\mathbf{R})\rangle=\int
\dd^{3n}\mathbf{r}\,\phi^*_i(\mathbf{r};\mathbf{R})\hat{O}\phi_j(\mathbf{r};\mathbf{R})$, with $\hat{O}$ a generic operator.
Note that in (\ref{Eqn:nuclear_WF_adiabatic}) we neglected the terms $\langle\phi_i(\mathbf{r};\mathbf{R})|\hat{T}|\phi_j(\mathbf{r};\mathbf{R})\rangle$, being of the order $1/M$ smaller than the kinetic energy of the 
electrons.\cite{Schwabl_07}

The NAC vectors $\mathbf{D}_{ij}(\mathbf{R})$ couple different adiabatic energy surfaces.
Generally the coupling terms in Eq.\ (\ref{Eqn:nuclear_WF_adiabatic}) are of order $1/\sqrt{M}$ smaller than the electronic energy,\cite{Schwabl_07} and therefore they can be safely neglected. In this case the adiabatic nuclear wave functions are evolved independently, i.e., their normalizations --- which give the adiabatic populations --- are constants of motion. The Born-Oppenheimer approximation is based on this decoupling.
However,
as the gap between two PESs narrows, the
NAC vectors become large, as it can be seen from the following expression:\cite{WC04}
\begin{equation}
\mathbf{D}_{ij}(\mathbf{R}) =
\label{Eqn:D_gap}
 -\frac{1}{M}\frac{\langle\phi_i(\mathbf{r};\mathbf{R})|(\nabla_{\mathbf{R}}\hat{H}^\mathrm{el})|\phi_j(\mathbf{r};\mathbf{R})\rangle}{E_i(\mathbf{R})-E_j(\mathbf{R})}
\;,
\end{equation}
where $E_i(\mathbf{R})-E_j(\mathbf{R})\neq0$ and the numerator remains finite.
This allows mixing between eigenstates for large enough nuclear velocities.
TSH then provides a convenient approximation of
the nonadiabatic molecular dynamics, i.e., of the electronic transitions that can occur along
with the nuclear motion.
\section{Beyond the Born-Oppenheimer approximation\label{sec:Tully}}
\subsection{Time-dependent perturbation theory}

A statistical reduction of a correlated electron-nuclear dynamics
into occasional, independent, PES
hopping is possible only if one transition (hop) is fully completed
before the next can take place. This requires that the nonadiabatic
coupling terms $\mathbf{D}_{ij}(\mathbf{R})$, although not
negligible, are small enough to be considered as a perturbation
during a short time interval $\Delta t$, as typically assumed in
analogous analyses, see, e.g., Refs.\ \onlinecite{Herman_JCP95,Herman_CP01,Pechukas_JCP72}.
In such a case, we can separate the Hamiltonian of the system in an
unperturbed part $\hat{H}_{{0,i}}=\hat{T}+E_i(\mathbf{R})$ and a
perturbation part
$- \I \sum_j\mathbf{D}_{ij}(\mathbf{R})\cdot\hat{\mathbf{P}}$. The aim is
to find an approximate solution of the time-dependent Schr\"odinger
equation (\ref{Eqn:nuclear_WF_adiabatic}) according to the
time-dependent perturbation theory.  In order to achieve this aim, it
is convenient to work in the interaction picture \cite{fetterwal},
where
\begin{eqnarray}
 \chi_{{i,I}}&=& e^{\I\hat{H}_{{0,i}}t}\chi_{{i}}\nonumber\quad ,\\
\I\partial_t\chi_{i,I}& =& \sum_{j}\hat{W}_{ij,I}\chi_{j,I}\nonumber\quad,\\ 
\hat{W}_{ij,I}&=& - \I\, e^{\I\hat{H}_{0,i}t} \left[
  \mathbf{D}_{ij}(\mathbf{R})\cdot\hat{\mathbf{P}} \right]
e^{-\I\hat{H}_{0,j}t}\nonumber\quad.  
\end{eqnarray}

Note that in the adiabatic representation no
time-ordering is needed in the definition of $\hat{W}_{ij}$.
Therefore the solution of the equation of motion for $\chi_{i,I}$, between 
initial time zero and final time $t$, at
first order in the perturbation $\sum_{j}\hat{W}_{ij,I}$ reads
\begin{eqnarray}
 \chi_{i,I}(\mathbf{R};t)
  &\approx&
 \chi_i(\mathbf{R};0)\nonumber\\
 &&
  -
  \sum_{j}\int_0^t \dd t'\,
   e^{\I\hat{H}_{0,i}t'}
  \mathbf{D}_{ij}(\mathbf{R})\cdot\hat{\mathbf{P}} \,
  e^{-\I\hat{H}_{0,j}t'}\chi_{j}(\mathbf{R};0).
  \label{Eqn:wavefunction_2}
\end{eqnarray}

\subsection{Approximations through Gaussians\label{sec:deriveG}}

Eventually, ionic motion is treated classically while the
  computation of a hopping probability has to proceed in a quantum
  mechanical framework.  In order to establish the link between these
  two descriptions, we need a \textit{semi-classical} approximation for the
  wave functions $\chi$. Inspired by Heller's work,\cite{Heller} the initial state $\chi_j(\mathbf{R};0)$ is represented in terms of Gaussian wave packets
\begin{equation}
  \mathcal{G}_{\mathbf{R}_i\mathbf{P}_i\lambda_i}(\mathbf{R})
  =
  \left(\frac{\lambda_i}{\pi}\right)^{3N/4}
  \exp\left[\I \mathbf{P}_i\cdot\mathbf{R}
  -\frac{\lambda_i}{2}(\mathbf{R}-\mathbf{R}_i)^2\right]
  \quad,
\end{equation}
as
\begin{equation}
  \chi_j(\mathbf{R};0)
  =
  \alpha_j
  \mathcal{G}_{\mathbf{R}_{j0}\mathbf{P}_{j0}\lambda_j}(\mathbf{R}),
\label{eq:chiansatz}
 \end{equation}
with $\mathbf{R}_{j0}$, $\mathbf{P}_{j0}$, and $\alpha_j$ being the
average positions, average momenta, and amplitudes, respectively, of
the wave packet at $t=0$ (see Appendix \ref{app:Gaussians} for more
details on the Gaussian wave packets). All Gaussians are
  normalized to one. The (complex) coefficient $\alpha_j$ regulates
  the contribution from each Gaussian to the whole state. It 
  expresses the correlations
  accumulated in previous time steps. The propagator
$e^{-\I\hat{H}_{0,j}t'}$ in (\ref{Eqn:wavefunction_2}) then evolves
this wave packet from the initial time $t=0$ to a time $t'$.
 At this point
we make use of a \textit{semi-classical approximation} for the nuclei: in the spirit of the frozen Gaussian approximation 
proposed by Heller,\cite{Heller} we impose that,
for a short time interval, the
width of the Gaussian is fixed (``frozen'') and that the time evolution of the
parameters $\mathbf{R}_{j}(t)$ and $\mathbf{P}_{j}(t)$ is given by the solution of the classical equations of motion
for an effective nuclear potential given by the $j^{th}$ PES. Note that the use of frozen Gaussians is a common practice
both in numerical applications and formal developments in the field.\cite{Rossky_96,Neria,Subotnik_11,Rossky_97,Herman_CP84,Herman_CP01,Herman_JCP04,Herman_JCP06}
One can then use the following approximation:%
\begin{eqnarray}
 e^{-\I\hat{H}_{0,j}t'}\mathcal{G}_{\mathbf{R}_{j0}\mathbf{P}_{j0}\lambda_j}
  &\approx&
   e^{-\I(E_j+T_j)t'}\mathcal{G}_{\mathbf{R}_{j}(t')\mathbf{P}_{j}(t')\lambda_j},
   \label{Eqn:wp_limit}
 \end{eqnarray}
  with $T_j=
  \frac{\mathbf{P}_j^2}{2M}$, $  E_j
  =
  E_j(\mathbf{R}_j)$, and 
their sum $T_j+E_j$ being constant along the classical evolution.
In doing so, we are neglecting the term $-\int_0^{t'} \dd t \,
\dot{\mathbf{P}}_j(t)\cdot\mathbf{R}_j(t)$ in the quantum phase accumulated
during the time evolution. This approximation is  justified for small momentum
changes during short-time propagation, which is in line with our derivation.
In the following,  we will refer to this semi-classical limit as the \textit{wave packet limit}. 
Note that this limit is in the spirit of the short-time expansion to a semi-classical golden rule employed, 
e.g., in Refs.\ \onlinecite{Neria,Rossky_96,Rossky_97,Rossky_98}.

For the sake of simplicity, we use a multivariate Gaussian with the same (frozen) width $\lambda_j$ 
for all nuclear Cartesian coordinates. This is justified 
in the classical limit $\lambda_j\rightarrow \infty$, which will be taken at the end of our derivation. 
Moreover in the following we will drop the time dependence of the average positions and momenta, if not needed. At each time the Gaussian wave packets fulfill the completeness relation 
\begin{eqnarray}
 \delta^{3N}(\mathbf{R}-\mathbf{R}')=\int \dd^{3N}\mathbf{P}_k\int\dd^{3N}\mathbf{P}_{\ell}\mathcal{G}_{\mathbf{R}_i\mathbf{P}_k\lambda_i}(\mathbf{R})\,
 \mathcal{I}_{\mathbf{R}_i\lambda_i}^{-1}(\mathbf{P}_k,\mathbf{P}_{\ell})
  \mathcal{G}^*_{\mathbf{R}_i\mathbf{P}_{\ell}\lambda_i}(\mathbf{R'}),
\label{eq:complete}
\end{eqnarray}
where $\mathcal{I}_{\mathbf{R}_i\lambda_i}^{-1}$ is the inverse of the overlap
$\mathcal{I}_{\mathbf{R}_i\lambda_i}(\mathbf{P}_k,\mathbf{P}_{\ell})=
\langle\mathcal{G}_{\mathbf{R}_i\mathbf{P}_k\lambda_i}|
    \mathcal{G}_{\mathbf{R}_i\mathbf{P}_{\ell}\lambda_i}\rangle$ (see Appendix \ref{app:Gaussians} for details). Inserting (\ref{eq:chiansatz}) and (\ref{eq:complete}) in (\ref{Eqn:wavefunction_2}) 
and applying the wave packet limit (\ref{Eqn:wp_limit}) yields
\begin{eqnarray}
  \chi_{i,I}(\mathbf{R};t)
  &\approx& \chi_{i}(\mathbf{R};0) -
  \sum_{j } \alpha_j
  \int_0^t \dd t'
     \int \dd^{3N}\mathbf{P}_k\int \dd^{3N}\mathbf{P}_{\ell}\,\nonumber\\
&&\qquad  
  e^{\I({E_i}+T_k-E_j-T_j)t'}
  \mathcal{G}_{\mathbf{R}_{i0}\mathbf{P}_{k0}\lambda_i}(\mathbf{R})\mathcal{I}_{\mathbf{R}_i\lambda_i}^{-1}(\mathbf{P}_k,\mathbf{P}_{\ell})\nonumber
\\
  &&\qquad
  \langle\mathcal{G}_{\mathbf{R}_i\mathbf{P}_{\ell}\lambda_i}|
  \mathbf{D}_{ij}
  |\mathcal{G}_{\mathbf{R}_j\mathbf{P}_j\lambda_j}\rangle
 \cdot\mathbf{P}_j
  \quad,
  \label{Eqn:X_I_G}
\end{eqnarray}
where we applied the wave packet limit also to the $\hat{\mathbf{P}}$
operator in (\ref{Eqn:wavefunction_2}) allowing the identification
$\hat{\mathbf{P}}\equiv\mathbf{P}_j$. Note that, in Eq.\ (\ref{Eqn:X_I_G}),
$\mathcal{G}_{\mathbf{R}_{i0}\mathbf{P}_{k0}\lambda_i}=e^{-\I(E_i+T_k)t'}e^{\I\hat{H}_{0,i}t'}\mathcal{G}_{\mathbf{R}_i(t')\mathbf{P}_k(t')\lambda_i}$
implicitly depends on $t'$ as initial time of the backward evolution of $(\mathbf{R}_i(t),\mathbf{P}_k(t))$ from $t'$ to $t=0$.

We have now to decide how to deal with the NAC vectors
$\mathbf{D}_{ij}$. 
The adiabatic
basis is associated with strongly varying
$\mathbf{D}_{ij}(\mathbf{R})$. Therefore, we will consider the following Gaussian
distribution for the coupling vectors
\begin{equation}
\label{eq:ansatz2}
\mathbf{D}_{ij}(\mathbf{R})
   =\mathbf{D}^{(ij)}_{0}\,
      \exp\left[-\left(\mathbf{R}-\mathbf{R}^{(ij)}_{\mathrm{c}}\right)^\mathrm{T}
            \hat{\mu}^{(ij)}\left(\mathbf{R}-\mathbf{R}^{(ij)}_{\mathrm{c}}\right)\right]
\;,
\end{equation}%
where $\mathrm{T} $ denotes transposition, $\mathbf{D}_0^{(ij)}$ is a constant, and $\mathbf{R}^{(ij)}_{\mathrm{c}}$ the
position of the avoided crossing.
For notational convenience we will drop the superscript ``(\textit{ij})'' on the right-hand side of Eq.\ (\ref{eq:ansatz2}).
The Gaussian ``width'' $\hat{\mu}^{(ij)}$ is a rank 2 tensor, whose form will be discussed in Sec.\  \ref{sec:curl}. Ansatz (\ref{eq:ansatz2}) is in line with the avoided crossing model proposed,
e.g., in Ref.~\onlinecite{Tully_90}, and with the analysis of Ref.~\onlinecite{Herman_CP01} based on
a semi-classical propagator; moreover it allows the NACs to fulfill the curl condition,\cite{Baer}
at least in the case of a two-level system (2LS), as it is illustrated in Sec.\ \ref{sec:curl}.

The Gaussian form for $\mathbf{D}_{ij}$ allows an analytical evaluation of the transition matrix elements. Moreover, to keep contact with the TSH technique, we consider that transitions $j\rightarrow i$ at an avoided crossing produce again wave packets of about the same spatial width (i.e., $\lambda_i=\lambda_j=\lambda$) and same average position (i.e.,  $\mathbf{R}_j=\mathbf{R}_i$).
Therefore, using the folding relations of Gaussians and the inverse $\mathcal{I}_{\mathbf{R}_j\lambda}^{-1}(\mathbf{P}_k,\mathbf{P}_{\ell})$ given in Appendix \ref{app:Gaussians}, one obtains
\begin{eqnarray}
\label{eq:chi_intstep}
  \chi_{i,I}(\mathbf{R};t)
  &\approx&
  \chi_{i}(\mathbf{R};0) - \left(\frac{1}{4\pi}\right)^{3N/2}
  \sum_{j}\frac{\alpha_j}{\sqrt{\det\left(\hat{\mu}\right)}}\int_0^t \dd t'  \int \dd^{3N}\mathbf{P}_k
\nonumber\\
  &&\qquad
 e^{\I(E_i+T_k-E_j-T_j)t'} \mathcal{G}_{\mathbf{R}_{i0}\mathbf{P}_{k0}\lambda}(\mathbf{R}) \;
  e^{\I(\mathbf{P}_j-\mathbf{P}_k)\mathbf{R}_\mathrm{c}}
\nonumber\\
 &&\qquad
  \mathbf{D}_0\cdot\mathbf{P}_j \;
  \exp\left[
    -\frac{1}{4}(\mathbf{P}_j-\mathbf{P}_k)^\mathrm{T}\hat{\mu}^{-1}
                                          (\mathbf{P}_j-\mathbf{P}_k)\right]
  \quad.
\end{eqnarray}
In principle, one cannot move the exponentials
 out of the time integral because
the wave packet parameters --- being evolved according to the classical equations of motion
--- can display a non-trivial time dependence.
On the other hand, we eventually consider the classical limit $\lambda\rightarrow \infty$ of the previous equation. In this limit, one can consider the evolution of the wave packet parameters to be smooth over a time scale, $t$, large
enough to approximate the time-integral with a Dirac delta-function.
This is also justified in the proper classical limit because energy fluctuations are suppressed, as
in classical molecular dynamics the total energy is exactly conserved at each time-step.
The result then becomes
\begin{eqnarray}
   \chi_{i,I}(\mathbf{R};t)
   &\approx& 
   \chi_{i}(\mathbf{R};0) -
   \left(\frac{1}{4\pi}\right)^{3N/2}   \sum_{j}\frac{\alpha_j}{\sqrt{\det\left(\hat{\mu}\right)}}
 \int \dd^{3N}\mathbf{P}_k\, 
\nonumber\\
 && \qquad
  \mathcal{G}_{\mathbf{R}_{i0}\mathbf{P}_{k0}\lambda}(\mathbf{R}) e^{\I(\mathbf{P}_j-\mathbf{P}_k)\mathbf{R}_\mathrm{c}}\mathbf{D}_0\cdot\mathbf{P}_j
\nonumber\\
 &&\qquad
  \exp\left[
    -\frac{1}{4}(\mathbf{P}_j-\mathbf{P}_k)^\mathrm{T}\hat{\mu}^{-1}
                                          (\mathbf{P}_j-\mathbf{P}_k)\right]
   \delta \left( E_i+\frac{\mathbf{P}_k^2}{2M}-E_j-\frac{\mathbf{P}_j^2}{2M} \right).
  \label{eq:cijGauss}
  \end{eqnarray}
We can now take the limit $\lambda\rightarrow \infty$, which simply
localizes the Gaussian wave packet
$\mathcal{G}_{\mathbf{R}_{i0}\mathbf{P}_{k0}\lambda}(\mathbf{R})$
while leaving unchanged the rest of the expression.

\subsection{The curl condition for the NACs\label{sec:curl}}
The NAC vectors are known to satisfy the so-called curl
condition if they are not in the neighborhood of a conical intersection.\cite{Baer}
For an arbitrary number of electronic PESs,
the curl condition is nonlinear in the components
of the NAC vectors, and its analysis is beyond the scope of this article.
However, for a 2LS with real wave functions
the curl condition is linear, and reads
\begin{equation}
\label{eq:curl-2levels}
\frac{\partial D^{(12)}_\alpha(\mathbf{R})}{\partial R_{\beta}}
-
\frac{\partial D^{(12)}_\beta(\mathbf{R})}{\partial R_{\alpha}}
=0
\;,
\end{equation}
where $D_{\alpha}^{(12)}(\mathbf{R})$ are the components of the
single nonzero independent NAC vector of the system,
$\mathbf{D}^{(12)}(\mathbf{R})=-\mathbf{D}^{(21)}(\mathbf{R})$,
and the equation holds for all pairs of nuclear coordinates $(\alpha,\beta)$. 

Ansatz (\ref{eq:ansatz2}) is flexible enough to adapt to the
curl condition for a 2LS, Eq.\ (\ref{eq:curl-2levels}). First, $\hat{\mu}$ should be a semi-positive symmetric matrix in the nuclear coordinates, i.e., it should satisfy $\mu_{\alpha\beta}=\mu_{\beta\alpha}$
and $\mathbf{R}^\mathrm{T}\hat{\mu}\mathbf{R}\geq0$ for all $\mathbf{R}$.
The properties of such a matrix $\hat{\mu}$
guarantee that i) there is a change of nuclear coordinates associated to an orthonormal basis change (i.e., a rotation)
which transforms $\hat{\mu}$ into diagonal form, and that ii) some of its eigenvalues are positive, and some may be zero.
Directions with zero eigenvalues of $\hat{\mu}$ give rise to Dirac deltas in momentum space instead of finite-width Gaussians, and hence there is strict momentum conservation along these directions; eigen-directions with nonzero eigenvalues are described via Gaussians in momentum space, with the $\hat{\mu}$ restricted to this invertible subspace, and hence small changes of the nuclear momentum along these directions are allowed.

It is possible to prove (see Appendix~\ref{app:mu_proof}) that for such an \textit{ansatz} the
curl condition is satisfied for all nuclear configurations $\mathbf{R}$
if and only if
\begin{equation}
\hat{\mu}\propto\mathbf{D}_0\otimes\mathbf{D}_0,
\label{eq:mu_d0}
\end{equation}
with a positive proportionality constant.
Such a $\hat{\mu}$ has a single nonzero eigenvalue, which
corresponds to the direction of $\mathbf{D}_0$. In the following we will consider a $\hat{\mu}$ as given in (\ref{eq:mu_d0}) also for the general case of multiple PESs.

\subsection{Transition rates \label{ssec:transition_rates}}
From the final result (\ref{eq:cijGauss}) of perturbation theory in
the Gaussian wave packet approximation, we can derive the change in time of the electronic
population in the $i^{th}$ state, $||\chi_i(t)||^2-||\chi_i(0)||^2$, 
from which the electronic transition rates between an initial state $\chi_j$ and the final states $\chi_i$ are obtained as
\begin{eqnarray}
  W_{j\mathbf{P}_j\rightarrow i\mathbf{P}_{k}}
  &\propto&
   \Re(\alpha_i^* \alpha_j) \,
  \mathbf{D}_0\cdot\mathbf{P}_j \,
  \exp\left[
    -\frac{1}{4}(\mathbf{P}_j-\mathbf{P}_{k})^\mathrm{T}\hat{\mu}^{-1}
                                          (\mathbf{P}_j-\mathbf{P}_{k})\right]
\nonumber\\
&&  \delta\left(E_i+\frac{\mathbf{P}_{k}^2}{2M}-E_j-\frac{\mathbf{P}_j^2}{2M}\right)+O((\mathbf{D}_0\cdot\mathbf{P}_j)^2).
\label{eq:finalW}
\end{eqnarray}
Equation (\ref{eq:finalW}) is the central result of this paper: 
it describes hopping between an initial adiabatic energy surface $E_j$, along which nuclei
move with momenta $\mathbf{P}_j$, and a final adiabatic surface $E_i$, along
which nuclei move with rescaled momenta $\mathbf{P}_{k}$ in order to conserve the
total energy. This is precisely the framework common to the various TSH approaches. 

Besides recovering the essential features of the TSH algorithm, our derivation provides a better understanding of the underlying physics. In particular, the change $\mathbf{P}_j\rightarrow \mathbf{P}_{k}$ in the nuclear momenta occurs within a range set by $\hat{\mu}$, which is related to the spatial variation of the nonadiabatic coupling vector. A similar result can be found in Ref.\ \onlinecite{Herman_CP01}. We note that when considering
nearly constant $\mathbf{D}_{ij}$, i.e., when $\hat{\mu}\longrightarrow 0$, the exponential in Eq.\ (\ref{eq:finalW}) becomes 
$\delta^{3N}(\mathbf{P}_{k}-\mathbf{P}_j)$ and the
energy matching becomes $\delta\left(E_i-E_j\right)$ requiring a
strict level crossing. This is the case in a diabatic basis, as we will see
in Sec.~\ref{sec:Illustration}.

The allowed changes in momentum are aligned along the direction of 
 $\mathbf{D}_0$, i.e., the direction of the NAC vectors, as discussed in Sec.\ \ref{sec:curl}. This result supports hence the
widely used procedure of adjusting the nuclear velocities along the NAC vectors
and it is in line with previous findings in this direction.
\cite{Tully_IJQC91,Coker_JCP95,Herman_JCP84,Herman_JCP95,Herman_JCP06} Note that this result strongly relies on the \textit{ansatz} (\ref{eq:ansatz2}) and (\ref{eq:mu_d0}) for the NACs. This choice allows the NACs to satisfy the curl condition for a 2LS (with real wave functions), Eq.\ (\ref{eq:curl-2levels}), and in our derivation we assume the same form also for a general multi-level system.

Finally, we find a transition rate linear in the
coupling vector $\mathbf{D}_0$, typical of the standard surface hopping
algorithm.\cite{Tully_90} 
Crucial in getting this scaling is to assume
that the state $\chi_i$ is initially populated in
Eq.\ (\ref{Eqn:wavefunction_2}) which means a non-zero
  $\alpha_i$ in Eq.\ (\ref{eq:finalW}). This requires that the
  electronic correlations contained in the $\alpha_j$ coefficients
  were propagated coherently over the history of the process.

 If one requires, instead, that a full state reduction is performed at each time when
  evaluating the transition rates, then each coherent propagation
  starts from a pure single state~$j'$, i.e., $\alpha_{j}=\delta_{jj'}$,
  which removes the sum over $j$ in (\ref{eq:cijGauss}).  In this case
the transition rates to previously unoccupied levels $i\ne j'$ are quadratic in $\mathbf{D}_0$,
since the linear term drops from Eq.\ (\ref{eq:finalW}):
\begin{eqnarray}
  W_{j'\mathbf{P}_{j'}\rightarrow i\mathbf{P}_{k}}
  &\propto&
 | \mathbf{D}_0\cdot\mathbf{P}_{j'} |^2\,
  \exp\left[
    -\frac{1}{2}(\mathbf{P}_{j'}-\mathbf{P}_{k})^\mathrm{T}\hat{\mu}^{-1}
                                          (\mathbf{P}_{j'}-\mathbf{P}_{k})\right]
\nonumber\\
&&  \delta\left(E_i+\frac{\mathbf{P}_{k}^2}{2M}-E_{j'}-\frac{\mathbf{P}_{j'}^2}{2M}\right).
\end{eqnarray}%
These results are in line with the recent findings that Tully's
surface hopping gives the wrong scaling in the spin-boson model (the
correct scaling being quadratic in the coupling vector) and that this
is due to an incorrect description of decoherence in the standard
TSH.\cite{Subotnik_comm}

\section{Illustration in the Landau-Zener model\label{sec:Illustration}}

\begin{figure}[tbp]
 \begin{center}
 \includegraphics[width=7.5cm]{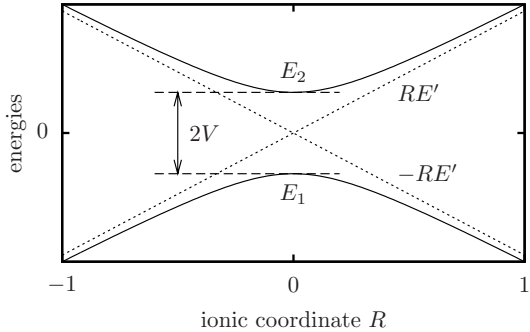}
 \end{center}
 \caption{\label{figure1}
Schematic plot of the diabatic energies
$\pm RE'$ (dotted lines) and adiabatic energies $E_{1,2}$ (solid lines) in the Landau-Zener model as functions
of the ionic distance $R$.}
\end{figure}

 The final formula (\ref{eq:finalW}) can be illustrated in the well known 
Landau-Zener model.\cite{Zener_32} The basics of the model are sketched
schematically in Fig.\ \ref{figure1}.  The nuclear degree of
freedom is described by one coordinate $R$.  The electronic degrees of freedom
are focused to a system of two levels $i=1,2$. The unperturbed system,
standing for the diabatic situation, has a linear
level crossing at $R=0$. The slope $E'$ is the first model
parameter. The strength of the interaction between the two diabatic levels is set by the coupling constant $V$,
which constitutes the second model parameter. The adiabatic representation is
obtained by solving the 2$\times$2 model Hamiltonian for fixed ionic
position. This yields the well known adiabatic PESs $E_{1,2}=\pm\sqrt{(RE')^2+V^2}$ as indicated in the
figure (solid lines). These are the PESs which enter the hopping formula
(\ref{eq:finalW}). The deviation from the diabatic energy levels is particularly strong around $R\approx{0}$,
which leads to a strongly varying nonadiabatic coupling
$\mathbf{D}_{ij}(\mathbf{R})$ as was assumed in
our derivation. It is a textbook exercise to work that out for the
Landau-Zener model. We find 
\begin{equation}
  \left|D_{12}(R)\right|
  \propto
  \frac{|E'/V|}{1+(E'/V)^2R^2}.
  \label{Eq:NAC_LZ}
  \end{equation}
This matrix element is strongly peaked at the avoided level crossing, i.e., around $R=0$, with
a characteristic $R$ width $\mu^{-1/2}\sim |V/E'|$. This translates to a typical
width of the momentum distribution of $|E'/V|$.  The example demonstrates that
some non-negligible momentum spread in the hopping is expected as we
usually encounter avoided crossings as modeled in the Landau-Zener model. The
overall strength of hopping matrix element is governed by the same parameter
combination $|E'/V|$ which determines the momentum width.  We thus find that
larger hopping probabilities are associated with larger momentum widths.

One can also try to generalize the Landau-Zener model to higher dimensions. A simplistic way to achieve this is to promote $E'$ to a vector, and, consequently, to interpret $RE'$ as a scalar product. In this case Eq.\ (\ref{Eq:NAC_LZ}) indicates that the direction of $D_{12}$ is along $E'$ and $\mu$ is proportional to the tensor product $E'\otimes E'$. This further supports \textit{ansatz} (\ref{eq:ansatz2}), with $\hat{\mu}\propto\mathbf{D}_0\otimes\mathbf{D}_0$, for the NACs, and our findings of Sec.\ \ref{ssec:transition_rates}.

\section{Conclusions \label{sec:conclusions}}

We propose a derivation of  the trajectory surface-hopping (TSH) technique based on a
semi-classical approximation to the nuclear dynamics in the spirit of a wave
packet limit.

The equations governing the electronic transition rates at a simple avoided level crossing display the essential features of  TSH algorithm and allows us to elucidate the underlying physics. We find a nonzero electronic transition rate at avoided crossings, which allows
for small changes in the nuclear momenta accounted for properly in the energy
matching. This justifies the rescaling of the classical velocities done in
practice after each hop. Moreover, we find that the physical source of the
width of allowed hops in momentum space is related to the speed of variation of
the nonadiabatic coupling elements in the adiabatic basis. In the classical
limit for the nuclei the derivation supports the rescaling of the momenta
along the nonadiabatic coupling vectors. This result strongly relies on the \textit{ansatz} employed for the nonadiabatic couplings (NACs). In our derivation we assume a multivariate Gaussian form for the NACs, which allows the NACs to fulfill of the so-called curl condition, at least in a two-level case.
We also find that the final
electronic transition rate is linear in the nonadiabatic coupling vectors, as
in the standard TSH algorithm, and that incorporating quantum decoherence makes
this scaling quadratic. 

Illustration through the Landau-Zener model supports our findings.

{\bfseries Acknowledgments:} Work supported by the project RE 322/10-1, Institut Universitaire de France, and Agence Nationale de la Recherche. LS acknowledges financial support from the Spanish MEC (FIS2011-65702-C02-01), ACI-Promociona (ACI2009-1036), Grupos Consolidados UPV/EHU del Gobierno Vasco (IT-319-07), the European Research Council Advanced Grant DYNamo (ERC-2010-AdG-Proposal No. 267374) and CRONOS (280879-2 CRONOS CP-FP7).

\appendix

\section{Gaussian wave packets \label{app:Gaussians}}

In this appendix, we collect a few general properties of the Gaussian
wave packets, for which most integrals are known analytically.  For
example, the folding of Gaussians in general obeys the simple rule
 \begin{eqnarray}
  \int \dd^{3N}\mathbf{R} && e^{\I\mathbf{P}\cdot \mathbf{R}}
  \,\exp\left(-\frac{(\mathbf{R}'-\mathbf{R})^2}{a}\right)
  \,\exp\left(-\frac{{(\mathbf{R}-\mathbf{R}'')}^2}{b}\right)\nonumber\\
& =&
  \left(\frac{ab\pi}{a+b}\right)^{3N/2}
  \,\exp\left(\I \mathbf{P}\frac{b\mathbf{R}'+a\mathbf{R}''}{a+b}\right)\nonumber\\
  &&\exp\left(-\frac{ab}{4(a+b)}\mathbf{P}^2-\frac{(\mathbf{R}'-\mathbf{R}'')^2}{a+b}\right)
  \quad.
\label{eq:foldG}
\end{eqnarray}

The basic multivariate isotropic Gaussian wave function reads
\begin{equation}
  \mathcal{G}_{\mathbf{R}_i\mathbf{P}_i\lambda}(\mathbf{R})
  =
  \left(\frac{\lambda}{\pi}\right)^{3N/4}
  \exp\left({\I} \mathbf{P}_i\cdot\mathbf{R}-\frac{\lambda}{2}(\mathbf{R}-\mathbf{R}_i)^2\right)
  \quad.
\label{eq:basG}
\end{equation}
where $\lambda$ controls the spatial width of the Gaussian.
In the present work we only consider overlaps between two Gaussian
wave packets with the same spatial widths and centers,
\begin{eqnarray}
  \mathcal{I}_{\mathbf{R}_i\lambda}(\mathbf{P}_i,\mathbf{P}_j)
  &=&
  \langle\mathcal{G}_{\mathbf{R}_i\mathbf{P}_i\lambda}|\mathcal{G}_{\mathbf{R}_i\mathbf{P}_j\lambda}\rangle\nonumber\\
&  =&
   e^{\I(\mathbf{P}_j-\mathbf{P}_i)\cdot\mathbf{R}_i}
    \exp\left(-\frac{(\mathbf{P}_i-\mathbf{P}_j)^2}{4\lambda}\right)
  \quad.
  \label{eq:IRoverlap}
\end{eqnarray}
The inverse of these overlaps, $\mathcal{I}^{-1}_{\mathbf{R}_i\lambda}(\mathbf{P}_k,\mathbf{P}_j)$
defined to satisfy
\begin{equation}
  \int \dd^{3N}\mathbf{P}_k\,\mathcal{I}_{\mathbf{R}_i\lambda}(\mathbf{P}_i,\mathbf{P}_k) \mathcal{I}^{-1}_{\mathbf{R}_i\lambda}(\mathbf{P}_k,\mathbf{P}_j)
  =
  \delta^{3N}(\mathbf{P}_i-\mathbf{P}_j)
   \quad,
\label{eq:definv}
\end{equation}
may be expressed as
\begin{eqnarray}
  \mathcal{I}^{-1}_{\mathbf{R}_i\lambda}(\mathbf{P}_i,\mathbf{P}_j)
 & =&
  \left(\frac{1}{4\pi\lambda}\right)^{3N/2}
  e^{-\I(\mathbf{P}_i-\mathbf{P}_j)\cdot\mathbf{R}_i}\nonumber\\
&&  \int\frac{\dd^{3N}\mathbf{Y}}{(2\pi)^{3N}}
  \exp\left(\lambda\mathbf{Y}^2\right)
  e^{\I\mathbf{Y}\cdot(\mathbf{P}_i-\mathbf{P}_j)}
  \quad.
\label{eq:inversR}
\end{eqnarray}
One can verify that such an inverse also satisfies the completeness relation (\ref{eq:complete}).

Details on more general multivariate Gaussians can be found, e.g., in Ref.\ \onlinecite{Riley}.

\section{Proof of the curl-consistency of $\hat{\mu}$ for a two-level system}\label{app:mu_proof}
In the basis where $\hat{\mu}$ is diagonal,
\begin{equation}
\mathbf{D}(\mathbf{R})
  = \mathbf{D}_0 \exp\left\{-\sum_{\alpha=1}^{3N} \mu_{\alpha}
                    \left[\left(\mathbf{R}-\mathbf{R}_{\mathrm{c}}\right)_\alpha\right]^2\right\}
\;,
\end{equation}
so
\begin{equation}
\label{eq:curl-2levels-ansatz2}
\frac{\partial D_{\alpha}(\mathbf{R})}{\partial R_{\beta}}
- \frac{\partial D_{\beta}(\mathbf{R})}{\partial R_{\alpha}}
=-2 \left[
       \mu_{\beta}\left(\mathbf{R}-\mathbf{R}_{\mathrm{c}}\right)_{\beta} D_{\alpha}(\mathbf{R})
     - \mu_{\alpha}\left(\mathbf{R}-\mathbf{R}_{\mathrm{c}}\right)_{\alpha} D_{\beta}(\mathbf{R})
    \right]
\;.
\end{equation}
If we enforce the curl condition for two levels at the point
$\mathbf{R}=\mathbf{R}_{\mathrm{c}}+c\hat{\mathbf{e}}_{\gamma}$,
where $(\mathbf{R}-\mathbf{R}_{\mathrm{c}})_{\alpha}= c\delta_{\alpha\gamma}$,
we get
\begin{equation}
\delta_{\beta\gamma} \mu_{\gamma} D_{0,\alpha}^{(12)} \exp\left(-\mu_{\gamma} c^2\right)
=
\delta_{\alpha\gamma} \mu_{\gamma} D_{0,\beta}^{(12)} \exp\left(-\mu_{\gamma} c^2\right)
\end{equation}
for all pairs of directions $(\alpha,\beta)$.
If we take $\beta=\gamma$, we have
\begin{equation}
\mu_{\gamma} D_{0,\alpha}^{(12)} = \delta_{\alpha\gamma} \mu_{\gamma}  D_{0,\gamma}^{(12)}
\end{equation}
for all components $\alpha$.
There are only two ways to satisfy these equations: either $\mu_{\gamma}=0$,
or $ D_{0,\alpha}^{(12)} = \delta_{\alpha\gamma} D_{0,\gamma}^{(12)}\;\forall\alpha$.
This implies that, if $\gamma$ is a direction such that $\mu_{\gamma}\neq 0$,
then, for all components $\alpha\neq\gamma$, $ D_{0,\alpha}^{(12)} = 0$.
Since, in order to have nonzero Gaussian NAC vectors, at least one component of $\mathbf{D}$ and one eigenvalue of $\hat{\mu}$
should be different from zero, there can only be one non-zero eigenvalue of $\hat{\mu}$,
and it will correspond to the same direction of the single non-zero
component of $\mathbf{D}_0^{(12)}$.
Therefore, in a base-independent expression,
$\hat{\mu} \propto \mathbf{D}_0 \otimes \mathbf{D}_0$, where the tensor product is defined
in terms of components as
$(a\otimes b)_{\alpha\beta}=a_{\alpha} b_{\beta}$,
and the proportionality constant must be a positive real number.
It is straightforward that this condition is not only necessary but sufficient, since
such a $\hat{\mu}$ will always satisfy Eq.\ (\ref{eq:curl-2levels-ansatz2}).

%


\begin{thebibliography}{35}%
\makeatletter
\providecommand \@ifxundefined [1]{%
 \@ifx{#1\undefined}
}%
\providecommand \@ifnum [1]{%
 \ifnum #1\expandafter \@firstoftwo
 \else \expandafter \@secondoftwo
 \fi
}%
\providecommand \@ifx [1]{%
 \ifx #1\expandafter \@firstoftwo
 \else \expandafter \@secondoftwo
 \fi
}%
\providecommand \natexlab [1]{#1}%
\providecommand \enquote  [1]{``#1''}%
\providecommand \bibnamefont  [1]{#1}%
\providecommand \bibfnamefont [1]{#1}%
\providecommand \citenamefont [1]{#1}%
\providecommand \href@noop [0]{\@secondoftwo}%
\providecommand \href [0]{\begingroup \@sanitize@url \@href}%
\providecommand \@href[1]{\@@startlink{#1}\@@href}%
\providecommand \@@href[1]{\endgroup#1\@@endlink}%
\providecommand \@sanitize@url [0]{\catcode `\\12\catcode `\$12\catcode
  `\&12\catcode `\#12\catcode `\^12\catcode `\_12\catcode `\%12\relax}%
\providecommand \@@startlink[1]{}%
\providecommand \@@endlink[0]{}%
\providecommand \url  [0]{\begingroup\@sanitize@url \@url }%
\providecommand \@url [1]{\endgroup\@href {#1}{\urlprefix }}%
\providecommand \urlprefix  [0]{URL }%
\providecommand \Eprint [0]{\href }%
\providecommand \doibase [0]{http://dx.doi.org/}%
\providecommand \selectlanguage [0]{\@gobble}%
\providecommand \bibinfo  [0]{\@secondoftwo}%
\providecommand \bibfield  [0]{\@secondoftwo}%
\providecommand \translation [1]{[#1]}%
\providecommand \BibitemOpen [0]{}%
\providecommand \bibitemStop [0]{}%
\providecommand \bibitemNoStop [0]{.\EOS\space}%
\providecommand \EOS [0]{\spacefactor3000\relax}%
\providecommand \BibitemShut  [1]{\csname bibitem#1\endcsname}%
\let\auto@bib@innerbib\@empty
\bibitem [{\citenamefont {Born}\ and\ \citenamefont
  {Huang}(2000)}]{Born-Huang}%
  \BibitemOpen
  \bibfield  {author} {\bibinfo {author} {\bibfnamefont {M.}~\bibnamefont
  {Born}}\ and\ \bibinfo {author} {\bibfnamefont {K.}~\bibnamefont {Huang}},\
  }\href@noop {} {\emph {\bibinfo {title} {Dynamical Theory of Crystal
  Lattices}}}\ (\bibinfo  {publisher} {Oxford University Press},\ \bibinfo
  {year} {2000})\BibitemShut {NoStop}%
\bibitem [{\citenamefont {Wodtke}, \citenamefont {Tully},\ and\ \citenamefont
  {Auerbach}(2004)}]{Wodtke}%
  \BibitemOpen
  \bibfield  {author} {\bibinfo {author} {\bibfnamefont {A.~M.}\ \bibnamefont
  {Wodtke}}, \bibinfo {author} {\bibfnamefont {J.~C.}\ \bibnamefont {Tully}}, \
  and\ \bibinfo {author} {\bibfnamefont {D.~J.}\ \bibnamefont {Auerbach}},\
  }\href {\doibase 10.1080/01442350500037521} {\bibfield  {journal} {\bibinfo
  {journal} {Int. Rev. in Phys. Chem.}\ }\textbf {\bibinfo {volume} {23}},\
  \bibinfo {pages} {513} (\bibinfo {year} {2004})}\BibitemShut {NoStop}%
\bibitem [{\citenamefont {Tully}\ and\ \citenamefont
  {Preston}(1971)}]{Tully_71}%
  \BibitemOpen
  \bibfield  {author} {\bibinfo {author} {\bibfnamefont {J.~C.}\ \bibnamefont
  {Tully}}\ and\ \bibinfo {author} {\bibfnamefont {R.~K.}\ \bibnamefont
  {Preston}},\ }\href {\doibase 10.1063/1.1675788} {\bibfield  {journal}
  {\bibinfo  {journal} {J. Chem. Phys.}\ }\textbf {\bibinfo {volume} {55}},\
  \bibinfo {pages} {562} (\bibinfo {year} {1971})}\BibitemShut {NoStop}%
\bibitem [{\citenamefont {Tully}(1990)}]{Tully_90}%
  \BibitemOpen
  \bibfield  {author} {\bibinfo {author} {\bibfnamefont {J.~C.}\ \bibnamefont
  {Tully}},\ }\href {\doibase 10.1063/1.459170} {\bibfield  {journal} {\bibinfo
   {journal} {J. Chem. Phys.}\ }\textbf {\bibinfo {volume} {93}},\ \bibinfo
  {pages} {1061} (\bibinfo {year} {1990})}\BibitemShut {NoStop}%
\bibitem [{\citenamefont {Hammes-Schiffer}\ and\ \citenamefont
  {Tully}(1994)}]{Tully_94}%
  \BibitemOpen
  \bibfield  {author} {\bibinfo {author} {\bibfnamefont {S.}~\bibnamefont
  {Hammes-Schiffer}}\ and\ \bibinfo {author} {\bibfnamefont {J.~C.}\
  \bibnamefont {Tully}},\ }\href {\doibase 10.1063/1.467455} {\bibfield
  {journal} {\bibinfo  {journal} {J. Chem. Phys.}\ }\textbf {\bibinfo {volume}
  {101}},\ \bibinfo {pages} {4657} (\bibinfo {year} {1994})}\BibitemShut
  {NoStop}%
\bibitem [{\citenamefont {Webster}\ \emph {et~al.}(1994)\citenamefont
  {Webster}, \citenamefont {Wang}, \citenamefont {Rossky},\ and\ \citenamefont
  {Friesner}}]{Webster_94}%
  \BibitemOpen
  \bibfield  {author} {\bibinfo {author} {\bibfnamefont {F.}~\bibnamefont
  {Webster}}, \bibinfo {author} {\bibfnamefont {E.~T.}\ \bibnamefont {Wang}},
  \bibinfo {author} {\bibfnamefont {P.~J.}\ \bibnamefont {Rossky}}, \ and\
  \bibinfo {author} {\bibfnamefont {R.~A.}\ \bibnamefont {Friesner}},\ }\href
  {\doibase 10.1063/1.467204} {\bibfield  {journal} {\bibinfo  {journal} {J.
  Chem. Phys.}\ }\textbf {\bibinfo {volume} {100}},\ \bibinfo {pages} {4835}
  (\bibinfo {year} {1994})}\BibitemShut {NoStop}%
\bibitem [{\citenamefont {Tully}(1998)}]{Tully_98}%
  \BibitemOpen
  \bibfield  {author} {\bibinfo {author} {\bibfnamefont {J.}~\bibnamefont
  {Tully}},\ }\href {\doibase 10.1039/a801824c} {\bibfield  {journal} {\bibinfo
   {journal} {Faraday Discuss.}\ }\textbf {\bibinfo {volume} {110}},\ \bibinfo
  {pages} {407} (\bibinfo {year} {1998})}\BibitemShut {NoStop}%
\bibitem [{\citenamefont {Prezhdo}\ and\ \citenamefont
  {Rossky}(1997{\natexlab{a}})}]{Prezhdo_97}%
  \BibitemOpen
  \bibfield  {author} {\bibinfo {author} {\bibfnamefont {O.}~\bibnamefont
  {Prezhdo}}\ and\ \bibinfo {author} {\bibfnamefont {P.}~\bibnamefont
  {Rossky}},\ }\href {\doibase 10.1063/1.474382} {\bibfield  {journal}
  {\bibinfo  {journal} {J. Chem. Phys.}\ }\textbf {\bibinfo {volume} {107}},\
  \bibinfo {pages} {825} (\bibinfo {year} {1997}{\natexlab{a}})}\BibitemShut
  {NoStop}%
\bibitem [{\citenamefont {Wong}\ and\ \citenamefont
  {Rossky}(2002{\natexlab{a}})}]{Wong_02}%
  \BibitemOpen
  \bibfield  {author} {\bibinfo {author} {\bibfnamefont {K.}~\bibnamefont
  {Wong}}\ and\ \bibinfo {author} {\bibfnamefont {P.}~\bibnamefont {Rossky}},\
  }\href {\doibase 10.1063/1.1468886} {\bibfield  {journal} {\bibinfo
  {journal} {J. Chem. Phys.}\ }\textbf {\bibinfo {volume} {116}},\ \bibinfo
  {pages} {8418} (\bibinfo {year} {2002}{\natexlab{a}})}\BibitemShut {NoStop}%
\bibitem [{\citenamefont {Wong}\ and\ \citenamefont
  {Rossky}(2002{\natexlab{b}})}]{Wong_02_2}%
  \BibitemOpen
  \bibfield  {author} {\bibinfo {author} {\bibfnamefont {K.}~\bibnamefont
  {Wong}}\ and\ \bibinfo {author} {\bibfnamefont {P.}~\bibnamefont {Rossky}},\
  }\href {\doibase 10.1063/1.1468887} {\bibfield  {journal} {\bibinfo
  {journal} {J. Chem. Phys.}\ }\textbf {\bibinfo {volume} {116}},\ \bibinfo
  {pages} {8429} (\bibinfo {year} {2002}{\natexlab{b}})}\BibitemShut {NoStop}%
\bibitem [{\citenamefont {Bedard-Hearn}, \citenamefont {Larsen},\ and\
  \citenamefont {Schwartz}(2005)}]{Schwartz_05}%
  \BibitemOpen
  \bibfield  {author} {\bibinfo {author} {\bibfnamefont {M.}~\bibnamefont
  {Bedard-Hearn}}, \bibinfo {author} {\bibfnamefont {R.}~\bibnamefont
  {Larsen}}, \ and\ \bibinfo {author} {\bibfnamefont {B.}~\bibnamefont
  {Schwartz}},\ }\href@noop {} {\bibfield  {journal} {\bibinfo  {journal} {J.
  Chem. Phys.}\ }\textbf {\bibinfo {volume} {123}},\ \bibinfo {pages} {234106}
  (\bibinfo {year} {2005})}\BibitemShut {NoStop}%
\bibitem [{\citenamefont {Subotnik}\ and\ \citenamefont
  {Shenvi}(2011)}]{Subotnik_11}%
  \BibitemOpen
  \bibfield  {author} {\bibinfo {author} {\bibfnamefont {J.~E.}\ \bibnamefont
  {Subotnik}}\ and\ \bibinfo {author} {\bibfnamefont {N.}~\bibnamefont
  {Shenvi}},\ }\href@noop {} {\bibfield  {journal} {\bibinfo  {journal} {J.
  Chem. Phys.}\ }\textbf {\bibinfo {volume} {134}},\ \bibinfo {pages} {024105}
  (\bibinfo {year} {2011})}\BibitemShut {NoStop}%
\bibitem [{\citenamefont {Subotnik}(2011)}]{Subotnik_11_2}%
  \BibitemOpen
  \bibfield  {author} {\bibinfo {author} {\bibfnamefont {J.~E.}\ \bibnamefont
  {Subotnik}},\ }\href {\doibase 10.1021/jp206557h} {\bibfield  {journal}
  {\bibinfo  {journal} {J. Phys. Chem. A}\ }\textbf {\bibinfo {volume} {115}},\
  \bibinfo {pages} {12083} (\bibinfo {year} {2011})}\BibitemShut {NoStop}%
\bibitem [{\citenamefont {Shenvi}, \citenamefont {Subotnik},\ and\
  \citenamefont {Yang}(2011)}]{Subotnik_11_3}%
  \BibitemOpen
  \bibfield  {author} {\bibinfo {author} {\bibfnamefont {N.}~\bibnamefont
  {Shenvi}}, \bibinfo {author} {\bibfnamefont {J.~E.}\ \bibnamefont
  {Subotnik}}, \ and\ \bibinfo {author} {\bibfnamefont {W.}~\bibnamefont
  {Yang}},\ }\href@noop {} {\bibfield  {journal} {\bibinfo  {journal} {J. Chem.
  Phys.}\ }\textbf {\bibinfo {volume} {134}},\ \bibinfo {pages} {144102}
  (\bibinfo {year} {2011})}\BibitemShut {NoStop}%
\bibitem [{\citenamefont {Hack}\ and\ \citenamefont
  {Truhlar}(2000)}]{Truhlar_00}%
  \BibitemOpen
  \bibfield  {author} {\bibinfo {author} {\bibfnamefont {M.}~\bibnamefont
  {Hack}}\ and\ \bibinfo {author} {\bibfnamefont {D.}~\bibnamefont {Truhlar}},\
  }\href {\doibase 10.1021/jp001629r} {\bibfield  {journal} {\bibinfo
  {journal} {J. Phys. Chem. A}\ }\textbf {\bibinfo {volume} {104}},\ \bibinfo
  {pages} {7917} (\bibinfo {year} {2000})}\BibitemShut {NoStop}%
\bibitem [{\citenamefont {Herman}(1984)}]{Herman_JCP84}%
  \BibitemOpen
  \bibfield  {author} {\bibinfo {author} {\bibfnamefont {M.~F.}\ \bibnamefont
  {Herman}},\ }\href {\doibase 10.1063/1.447708} {\bibfield  {journal}
  {\bibinfo  {journal} {J. Chem. Phys.}\ }\textbf {\bibinfo {volume} {81}},\
  \bibinfo {pages} {754} (\bibinfo {year} {1984})}\BibitemShut {NoStop}%
\bibitem [{\citenamefont {Herman}(1995)}]{Herman_JCP95}%
  \BibitemOpen
  \bibfield  {author} {\bibinfo {author} {\bibfnamefont {M.~F.}\ \bibnamefont
  {Herman}},\ }\href {\doibase 10.1063/1.470173} {\bibfield  {journal}
  {\bibinfo  {journal} {J. Chem. Phys.}\ }\textbf {\bibinfo {volume} {103}},\
  \bibinfo {pages} {8081} (\bibinfo {year} {1995})}\BibitemShut {NoStop}%
\bibitem [{\citenamefont {Wu}\ and\ \citenamefont
  {Herman}(2006)}]{Herman_JCP06}%
  \BibitemOpen
  \bibfield  {author} {\bibinfo {author} {\bibfnamefont {Y.}~\bibnamefont
  {Wu}}\ and\ \bibinfo {author} {\bibfnamefont {M.~F.}\ \bibnamefont
  {Herman}},\ }\href {\doibase 10.1063/1.2358352} {\bibfield  {journal}
  {\bibinfo  {journal} {J. Chem. Phys}\ }\textbf {\bibinfo {volume} {125}},\
  \bibinfo {eid} {154116} (\bibinfo {year} {2006})}\BibitemShut {NoStop}%
\bibitem [{\citenamefont {Schwabl}(2007)}]{Schwabl_07}%
  \BibitemOpen
  \bibfield  {author} {\bibinfo {author} {\bibfnamefont {F.}~\bibnamefont
  {Schwabl}},\ }\href {\doibase 10.1007/978-3-540-71933-5} {\emph {\bibinfo
  {title} {Quantum Mechanics}}},\ \bibinfo {edition} {4th}\ ed.\ (\bibinfo
  {publisher} {Springer},\ \bibinfo {year} {2007}),\ p.\ \bibinfo {pages} {273--275}
  \BibitemShut {NoStop}%
\bibitem [{\citenamefont {Worth}\ and\ \citenamefont
  {Cederbaum}(2004)}]{WC04}%
  \BibitemOpen
  \bibfield  {author} {\bibinfo {author} {\bibfnamefont {G.~A.}~\bibnamefont
  {Worth}}\ and\ \bibinfo {author} {\bibfnamefont {L.~S.}\ \bibnamefont
  {Cederbaum}},\ }\href {\doibase 10.1146/annurev.physchem.55.091602.094335} {\bibfield  {journal}
  {\bibinfo  {journal} {Annu. Rev. Phys. Chem.}\ }\textbf {\bibinfo {volume} {55}},\
  \bibinfo {pages} {127} (\bibinfo {year} {2004})}\BibitemShut {NoStop}%
\bibitem [{\citenamefont {Herman}(2001)}]{Herman_CP01}%
  \BibitemOpen
  \bibfield  {author} {\bibinfo {author} {\bibfnamefont {M.~F.}\ \bibnamefont
  {Herman}},\ }\href {\doibase 10.1016/S0301-0104(01)00479-7} {\bibfield
  {journal} {\bibinfo  {journal} {Chem. Phys.}\ }\textbf {\bibinfo {volume}
  {273}},\ \bibinfo {pages} {175 } (\bibinfo {year} {2001})}\BibitemShut
  {NoStop}%
\bibitem [{\citenamefont {Pechukas}\ and\ \citenamefont
  {Davis}(1972)}]{Pechukas_JCP72}%
  \BibitemOpen
  \bibfield  {author} {\bibinfo {author} {\bibfnamefont {P.}~\bibnamefont
  {Pechukas}}\ and\ \bibinfo {author} {\bibfnamefont {J.~P.}\ \bibnamefont
  {Davis}},\ }\href {\doibase 10.1063/1.1676976} {\bibfield  {journal}
  {\bibinfo  {journal} {J. Chem. Phys}\ }\textbf {\bibinfo {volume} {56}},\
  \bibinfo {pages} {4970} (\bibinfo {year} {1972})}\BibitemShut {NoStop}%
\bibitem [{\citenamefont {Fetter}\ and\ \citenamefont
  {Walecka}(2003)}]{fetterwal}%
  \BibitemOpen
  \bibfield  {author} {\bibinfo {author} {\bibfnamefont {A.}~\bibnamefont
  {Fetter}}\ and\ \bibinfo {author} {\bibfnamefont {J.~D.}\ \bibnamefont
  {Walecka}},\ }\href@noop {} {\emph {\bibinfo {title} {Quantum Theory of
  Many-Particle Systems}}}\ (\bibinfo  {publisher} {Dover},\ \bibinfo {year}
  {2003})\BibitemShut {NoStop}%
\bibitem [{\citenamefont {Heller}(1981)}]{Heller}%
  \BibitemOpen
  \bibfield  {author} {\bibinfo {author} {\bibfnamefont {E.~J.}\ \bibnamefont
  {Heller}},\ }\href {\doibase 10.1063/1.442382} {\bibfield  {journal}
  {\bibinfo  {journal} {J. Chem. Phys.}\ }\textbf {\bibinfo {volume} {75}},\
  \bibinfo {pages} {2923} (\bibinfo {year} {1981})}\BibitemShut {NoStop}%
\bibitem [{\citenamefont {Schwartz}\ \emph {et~al.}(1996)\citenamefont
  {Schwartz}, \citenamefont {Bittner}, \citenamefont {Prezhdo},\ and\
  \citenamefont {Rossky}}]{Rossky_96}%
  \BibitemOpen
  \bibfield  {author} {\bibinfo {author} {\bibfnamefont {B.~J.}\ \bibnamefont
  {Schwartz}}, \bibinfo {author} {\bibfnamefont {E.~R.}\ \bibnamefont
  {Bittner}}, \bibinfo {author} {\bibfnamefont {O.~V.}\ \bibnamefont
  {Prezhdo}}, \ and\ \bibinfo {author} {\bibfnamefont {P.~J.}\ \bibnamefont
  {Rossky}},\ }\href {\doibase 10.1063/1.471326} {\bibfield  {journal}
  {\bibinfo  {journal} {J. Chem. Phys.}\ }\textbf {\bibinfo {volume} {104}},\
  \bibinfo {pages} {5942} (\bibinfo {year} {1996})}\BibitemShut {NoStop}%
\bibitem [{\citenamefont {Neria}\ and\ \citenamefont {Nitzan}(1993)}]{Neria}%
  \BibitemOpen
  \bibfield  {author} {\bibinfo {author} {\bibfnamefont {E.}~\bibnamefont
  {Neria}}\ and\ \bibinfo {author} {\bibfnamefont {A.}~\bibnamefont {Nitzan}},\
  }\href {\doibase 10.1063/1.465409} {\bibfield  {journal} {\bibinfo  {journal}
  {J. Chem. Phys.}\ }\textbf {\bibinfo {volume} {99}},\ \bibinfo {pages} {1109}
  (\bibinfo {year} {1993})}\BibitemShut {NoStop}%
\bibitem [{\citenamefont {Prezhdo}\ and\ \citenamefont
  {Rossky}(1997{\natexlab{b}})}]{Rossky_97}%
  \BibitemOpen
  \bibfield  {author} {\bibinfo {author} {\bibfnamefont {O.~V.}\ \bibnamefont
  {Prezhdo}}\ and\ \bibinfo {author} {\bibfnamefont {P.~J.}\ \bibnamefont
  {Rossky}},\ }\href {\doibase 10.1063/1.474312} {\bibfield  {journal}
  {\bibinfo  {journal} {J. Chem. Phys.}\ }\textbf {\bibinfo {volume} {107}},\
  \bibinfo {pages} {5863} (\bibinfo {year} {1997}{\natexlab{b}})}\BibitemShut
  {NoStop}%
\bibitem [{\citenamefont {Herman}\ and\ \citenamefont
  {Kluk}(1984)}]{Herman_CP84}%
  \BibitemOpen
  \bibfield  {author} {\bibinfo {author} {\bibfnamefont {M.~F.}\ \bibnamefont
  {Herman}}\ and\ \bibinfo {author} {\bibfnamefont {E.}~\bibnamefont {Kluk}},\
  }\href {\doibase 10.1016/0301-0104(84)80039-7} {\bibfield  {journal}
  {\bibinfo  {journal} {Chem. Phys.}\ }\textbf {\bibinfo {volume} {91}},\
  \bibinfo {pages} {27 } (\bibinfo {year} {1984})}\BibitemShut {NoStop}%
\bibitem [{\citenamefont {Herman}, \citenamefont {Akramine},\ and\
  \citenamefont {Moody}(2004)}]{Herman_JCP04}%
  \BibitemOpen
  \bibfield  {author} {\bibinfo {author} {\bibfnamefont {M.~F.}\ \bibnamefont
  {Herman}}, \bibinfo {author} {\bibfnamefont {O.~E.}\ \bibnamefont
  {Akramine}}, \ and\ \bibinfo {author} {\bibfnamefont {M.~P.}\ \bibnamefont
  {Moody}},\ }\href {\doibase 10.1063/1.1687313} {\bibfield  {journal}
  {\bibinfo  {journal} {J. Chem. Phys}\ }\textbf {\bibinfo {volume} {120}},\
  \bibinfo {pages} {7383} (\bibinfo {year} {2004})}\BibitemShut {NoStop}%
\bibitem [{\citenamefont {Prezhdo}\ and\ \citenamefont
  {Rossky}(1998)}]{Rossky_98}%
  \BibitemOpen
  \bibfield  {author} {\bibinfo {author} {\bibfnamefont {O.~V.}\ \bibnamefont
  {Prezhdo}}\ and\ \bibinfo {author} {\bibfnamefont {P.~J.}\ \bibnamefont
  {Rossky}},\ }\href {\doibase 10.1103/PhysRevLett.81.5294} {\bibfield
  {journal} {\bibinfo  {journal} {Phys. Rev. Lett.}\ }\textbf {\bibinfo
  {volume} {81}},\ \bibinfo {pages} {5294} (\bibinfo {year}
  {1998})}\BibitemShut {NoStop}%
\bibitem [{\citenamefont {Baer}(2006)}]{Baer}%
  \BibitemOpen
  \bibfield  {author} {\bibinfo {author} {\bibfnamefont {M.}~\bibnamefont
  {Baer}},\ }\href@noop {} {\emph {\bibinfo {title} {Beyond
  Born-Oppenheimer}}}\ (\bibinfo  {publisher} {John Wiley $\&$ Sons},\ \bibinfo
  {year} {2006})\BibitemShut {NoStop}%
\bibitem [{\citenamefont {Tully}(1991)}]{Tully_IJQC91}%
  \BibitemOpen
  \bibfield  {author} {\bibinfo {author} {\bibfnamefont {J.~C.}\ \bibnamefont
  {Tully}},\ }\href {\doibase 10.1002/qua.560400830} {\bibfield  {journal}
  {\bibinfo  {journal} {Int. J. Quant. Chem.}\ }\textbf {\bibinfo {volume}
  {40}},\ \bibinfo {pages} {299} (\bibinfo {year} {1991})}\BibitemShut
  {NoStop}%
\bibitem [{\citenamefont {Coker}\ and\ \citenamefont
  {Xiao}(1995)}]{Coker_JCP95}%
  \BibitemOpen
  \bibfield  {author} {\bibinfo {author} {\bibfnamefont {D.~F.}\ \bibnamefont
  {Coker}}\ and\ \bibinfo {author} {\bibfnamefont {L.}~\bibnamefont {Xiao}},\
  }\href {\doibase 10.1063/1.469428} {\bibfield  {journal} {\bibinfo  {journal}
  {J. Chem. Phys}\ }\textbf {\bibinfo {volume} {102}},\ \bibinfo {pages} {496}
  (\bibinfo {year} {1995})}\BibitemShut {NoStop}%
\bibitem [{\citenamefont {Landry}\ and\ \citenamefont
  {Subotnik}(2011)}]{Subotnik_comm}%
  \BibitemOpen
  \bibfield  {author} {\bibinfo {author} {\bibfnamefont {B.~R.}\ \bibnamefont
  {Landry}}\ and\ \bibinfo {author} {\bibfnamefont {J.~E.}\ \bibnamefont
  {Subotnik}},\ }\href {\doibase 10.1063/1.3663870} {\bibfield  {journal}
  {\bibinfo  {journal} {J. Chem. Phys.}\ }\textbf {\bibinfo {volume} {135}},\
  \bibinfo {eid} {191101} (\bibinfo {year} {2011})}\BibitemShut {NoStop}%
\bibitem [{\citenamefont {Zener}(1932)}]{Zener_32}%
  \BibitemOpen
  \bibfield  {author} {\bibinfo {author} {\bibfnamefont {C.}~\bibnamefont
  {Zener}},\ }\href {\doibase 10.1098/rspa.1932.0165} {\bibfield  {journal}
  {\bibinfo  {journal} {Proc. Royal Soc. A}\ }\textbf {\bibinfo {volume}
  {137}},\ \bibinfo {pages} {696} (\bibinfo {year} {1932})}\BibitemShut
  {NoStop}%
\bibitem [{\citenamefont {Riley}, \citenamefont {Hobson},\ and\ \citenamefont
  {Bence}(2006)}]{Riley}%
  \BibitemOpen
  \bibfield  {author} {\bibinfo {author} {\bibfnamefont {K.~F.}\ \bibnamefont
  {Riley}}, \bibinfo {author} {\bibfnamefont {M.~P.}\ \bibnamefont {Hobson}}, \
  and\ \bibinfo {author} {\bibfnamefont {S.~J.}\ \bibnamefont {Bence}},\
  }\href@noop {} {\emph {\bibinfo {title} {Mathematical Methods for Physics and
  Engineering}}},\ \bibinfo {edition} {3rd}\ ed.\ (\bibinfo  {publisher}
  {Cambridge University Press},\ \bibinfo {year} {2006})\BibitemShut {NoStop}%
\end{thebibliography}
\end{document}